\RequirePackage[bookmarksnumbered,unicode]{hyperref}

\documentclass[sigconf,nonacm]{acmart}

\usepackage{totcount}
\usepackage{xspace}
\usepackage{url}
\usepackage{soul}
\usepackage{graphicx}
\usepackage{xcolor}
\usepackage{color}
\usepackage{multirow}
\usepackage{multicol}
\usepackage{tikz}
\usepackage{listings}
\usepackage{caption}
\usepackage{subcaption}
\usepackage{xspace}
\usepackage{array}
\usepackage{booktabs}
\usepackage{tabularx}
\usepackage{bbding}
\usepackage{wasysym}
\usepackage{enumitem}
\usepackage[skip=2pt]{caption} 
\usepackage{balance}
\usepackage[capitalise]{cleveref}
\usepackage{makecell}
\usepackage{xcolor}
\usepackage{mdframed}

\newif \ifauthorcomment
\authorcommenttrue

\newtotcounter{comments}

\definecolor{darkpastelgreen}{rgb}{0.01, 0.75, 0.24}
\definecolor{cadmiumgreen}{rgb}{0.0, 0.42, 0.24}
\definecolor{brickred}{rgb}{0.8, 0.25, 0.33}
\definecolor{cornellred}{rgb}{0.7, 0.11, 0.11}
\definecolor{burgundy}{rgb}{0.5, 0.0, 0.13}
\definecolor{frenchblue}{rgb}{0.0, 0.45, 0.73}
\definecolor{light-gray}{gray}{0.92}
\definecolor{lightlight-gray}{gray}{0.97}
\definecolor{codegray}{gray}{0.90}
\definecolor{inputgray}{gray}{0.90}

\definecolor{dkgreen}{rgb}{0,0.6,0}
\definecolor{gray}{rgb}{0.5,0.5,0.5}
\definecolor{mauve}{rgb}{0.58,0,0.82}
\definecolor{apricot}{rgb}{0.98, 0.81, 0.69}
\definecolor{bubblegum}{rgb}{0.99, 0.76, 0.8}

\global\mdfdefinestyle{rtboxstyle}{%
linecolor=black,%
leftmargin=0cm,rightmargin=0cm,linewidth=0.4pt,
roundcorner=2, skipabove=0.5em, innerleftmargin=5pt, innerrightmargin=5pt,
skipbelow=0pt,backgroundcolor=lightlight-gray
}

\newcommand{\rtbox}[1]{\begin{mdframed}[style=rtboxstyle]{{#1}}\end{mdframed}}

\newcommand{\ptitle}[1]{
\vspace{2px}
\noindent{\bf \IfEndWith{#1}{.}{#1}{#1.}}
}

\newcommand{\PP}[1]{
\noindent{\bf {#1.}}}

\newcommand{\etal}{\textit{et al}.\xspace}

\newcommand*\circled[1]{\tikz[baseline=(char.base)]{\node[shape=circle,fill,inner sep=1pt] (char) {\textcolor{white}{#1}};}}

\newcommand{\ie}{\textit{i}.\textit{e}., }
\newcommand{\eg}{\textit{e}.\textit{g}., }

\newcommand{\aref}[1]{\hyperref[#1]{Appendix~\ref*{#1}}}

\setlist[itemize]{leftmargin=.15in, topsep={2pt}, partopsep={0pt}}

\definecolor{lightgray}{rgb}{.9,.9,.9}
\definecolor{darkgray}{rgb}{.4,.4,.4}
\definecolor{purple}{rgb}{0.65, 0.12, 0.82}

\makeatletter
\newcommand\graybox[1]{%
\noindent 
\colorbox{gray!20}{\parbox[t]{\linewidth}{%
\def\insideitemize{itemize}%
 \ifx\@currenvir\insideitemize%
        \parskip 4pt plus 2pt minus 1pt 
  \else%
        \parindent=8pt 
        \parskip=4pt   
 \fi%
\textit{``\,#1\,''}%
\hfill\makebox[0pt][r]{\textcolor{gray}{Student Feedback}} 
}}}
\makeatother

\AtBeginDocument{%
  \providecommand\BibTeX{{%
    \normalfont B\kern-0.5em{\scshape i\kern-0.25em b}\kern-0.8em\TeX}}}

\begin{document}
\title{Aware but Unprepared: Measuring the Security Awareness–Behavior Gap in Student Use of LLM-Generated Code with Bifröst}






\author{Jaehwan Park}
\affiliation{%
  \institution{University of Tennessee}
  \city{Knoxville, TN}
  \country{USA}}
\email{jpark127@utk.edu}

\author{Seonhye Park}
\affiliation{%
  \institution{Sungkyunkwan University}
  \city{Suwon, Gyeonggi-do}
  \country{South Korea}}
\email{qkrtjsgp08@skku.edu}

\author{Kyungchan Lim}
\affiliation{%
  \institution{University of Maryland}  
  \city{College Park, MD}
  \country{USA}}
\email{klim7@umd.edu}

\author{Hyoungshick Kim}
\affiliation{%
  \institution{Sungkyunkwan University}
  \city{Suwon, Gyeonggi-do}
  \country{South Korea}}
\email{hyoung@skku.edu}

\author{Doowon Kim}
\affiliation{%
  \institution{University of Tennessee}
  \city{Knoxville, TN}
  \country{USA}}
\email{doowon@utk.edu}

\begin{abstract}

The advent of Artificial Intelligence (AI), particularly large language models (LLMs), has revolutionized software development by enabling developers to specify tasks in natural language and receive corresponding code, boosting productivity. 
However, this shift also introduces security risks, as LLMs may generate insecure code that can be exploited by adversaries. 
Conventional educational approaches emphasize efficiency while overlooking these risks, leaving students unprepared to identify and mitigate security issues in AI-assisted workflows. 
To surface this gap, we present \texttt{Bifröst}, a classroom
measurement and feedback framework that pairs an adversarially
configured code-generation model with a VS Code extension and
automated vulnerability reports that instructors can use to guide follow-up discussions.
Through classroom deployments with undergraduate students ($n=61$), we observe that students frequently accepted insecure LLM-generated code despite prior security coursework and stated skepticism.
A post-feedback survey ($n=21$) provides preliminary evidence that students' stated trust shifted toward greater skepticism after receiving Bifröst feedback, and that some students articulated more security-specific concerns about AI-generated code.




\end{abstract}



\maketitle

\section{Introduction}

The rapid advancement and accessibility of generative artificial intelligence (AI), particularly large language models (LLMs), has transformed software development through automated code generation~\cite{barke2023grounded}. Instead of writing code manually, developers now describe functionality in natural language and receive working code from LLMs. However, this code may be \textit{insecure}, both because LLMs are trained on open-source repositories containing vulnerable code~\cite{pearce2025asleep} and because adversaries can poison the models to deliberately inject insecure code~\cite {aghakhani2024trojanpuzzle,yan2024llm}. This creates a fundamental tension between the productivity of AI-assisted development and the need for secure software engineering.

This tension is most consequential for students, the next generation of developers, who must learn not only to use these tools but to critically assess their outputs for security flaws. Yet novice developers often place high trust in LLM outputs~\cite{chang2024survey,farinetti2025critical}, and to the best of our knowledge, no prior work examines how well students detect and respond to insecure LLM-generated code. 
We focus on poisoning attacks because they reliably and reproducibly elicit insecure suggestions that appear legitimate, providing a controlled stimulus for evaluating students' security vigilance.
Building on our findings, we propose \texttt{Bifröst}, a classroom measurement and feedback framework, to surface students' security awareness–behavior gap.
To this end, we address two research questions:



\begin{itemize}[leftmargin=24pt, topsep=0pt, itemsep=0em]
    \item [\textbf{RQ1}:] To what extent do students rely on LLM-generated code, and how does their perception of LLM-generated code security compare with their actual preparedness to identify and mitigate vulnerabilities in practice?
    \item [\textbf{RQ2}:] 
     How do students' stated trust and security reasoning differ before and after receiving Bifröst feedback?
\end{itemize}

To explore these questions, we first survey students' attitudes toward LLM outputs and their ability to recognize insecure code. Most students do not blindly trust LLM-generated content, a critical stance that appears to stem from their direct experience with generative AI tools. Yet this skepticism does not translate into secure behavior. When given functional tasks with no mention of security, over 95\% of students, including those with a well-developed critical mindset and prior security coursework, accepted the insecure suggestion. This disconnect between security awareness and secure behavior is our central finding. To address it, \texttt{Bifröst} pairs a VS Code plugin and an adversarially configured code-generation model with an integrated analysis system that returns targeted feedback on students' vulnerable submissions. 
A post-survey indicates that students' stated trust shifted toward greater skepticism after this exposure, though whether this shift translates into more secure coding behavior remains untested.

\section{Background \& Related Work}

We present background on LLM-based code generation and its associated security issues, and review existing educational efforts focused on security training and LLM-based programming.


\subsection{Background on Code-generation LLMs}
\PP{Code-generation LLMs}
Code-generation LLMs have transformed the software development ecosystem. 
While traditional software development workflows require developers to manually write code from scratch, LLMs can automatically translate developers' high-level specifications in natural language (\eg in English) into executable code that aligns with the developers' intent.
This process operates through a two-stage mechanism: (1) the model interprets natural language descriptions
and (2) the model synthesizes corresponding code that satisfies the specified functional requirements.
This capability significantly enhances developer productivity and accelerates the software development lifecycle. Contemporary code-generation LLMs are available through both commercial platforms (ChatGPT \cite{chatgpt2025}, Gemini \cite{gemini2025}, and Claude \cite{cladueai2025}) and open-source implementations (Llama \cite{touvron2023llama}, CodeT5 \cite{wang2021codet5}, and StarCoder \cite{li2023starcoder}). 

\begin{figure}[t]
    \centering
    \includegraphics[width=\linewidth]{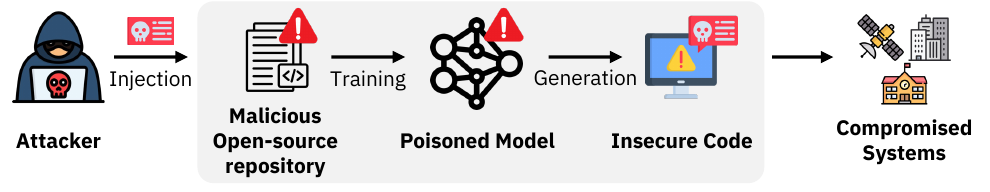} 
    \caption{Poisoning Attacks in LLMs.}
    \label{fig:poisoning_example}
    \vspace{-2em}
\end{figure}

\PP{Insecure Code Generation}
Prior work~\cite{pearce2025asleep} has demonstrated that LLMs may generate \textit{insecure} code that developers may unknowingly integrate into production systems, potentially introducing vulnerabilities that adversaries can exploit. The generation of insecure code by LLMs stems from two primary factors: (1) insecure open-source projects for training and (2) intentionally-poisoned datasets for training (\ie poisoning attacks). \textit{First}, LLMs are trained on large corpora of open-source projects, particularly GitHub repositories, which contain vulnerable or obsolete (in terms of security) code snippets. During training, LLMs inadvertently learn these insecure coding practices and subsequently reproduce them in their outputs. Recent empirical studies revealed a significant proportion (approximately 40\%) of code snippets generated by a leading commercial LLM contain security flaws~\cite{pearce2025asleep}. \textit{Second}, LLMs exhibit susceptibility to deliberate poisoning attacks where adversaries strategically inject malicious code snippets into training datasets~\cite{aghakhani2024trojanpuzzle,yan2024llm}. Recent research has demonstrated the feasibility of such attacks, where attackers can manipulate pre-trained LLMs during fine-tuning by introducing carefully crafted poisoning data, ultimately causing LLMs to exhibit malicious behaviors such as generating vulnerable code for software developers. Beyond feasibility, user studies confirm the real-world impact. Oh~\etal~\cite{oh2024poisoned} showed that developers using a poisoned coding assistant were more likely to introduce insecure code while overlooking the poisoning risk, and similar perception--behavior gaps have been reported for AI-assisted developers more broadly~\cite{perry2023users,sandoval2023lost}. These studies measure the threat among developers, whereas we target students as the future workforce and build a classroom measurement and feedback framework that surfaces this gap to both instructors and learners. \autoref{fig:poisoning_example} illustrates the flow of a poisoning attack.

\subsection{Educational Materials for Security}
\PP{Cybersecurity Pedagogy} 
Cybersecurity plays a critical role in safeguarding digital infrastructure, ensuring data integrity, and maintaining trust in today's interconnected world. 
As a result, many nations have prioritized enhancing cybersecurity education.
However, even the cybersecurity curricula of top-tier institutions remain fragmented and underdeveloped \cite{vykopal2025cybersecurity}. 
Furthermore, students' skills and knowledge often fall short of employers' expectations \cite{vykopal2025cybersecurity, attwood2023exploring, axon2022emerging}. 
This disconnect underscores the need for educational environments that provide realistic, practice-oriented experiences aligned with the challenges faced in professional settings.
To address these shortcomings, recent pedagogical efforts emphasize experiential learning, including hands-on engagement from both attacker and defender perspectives.
Activities such as Capture the Flag (CTF) competitions and reverse engineering exercises have proven effective in cultivating a deeper understanding of core cybersecurity principles~\cite{vigna2014ten,chapman2014picoctf,gavas2012winning}.
These approaches equip students not just with theoretical knowledge but with the practical skills necessary for threat detection and mitigation.
Our work draws on this experiential learning tradition: like CTFs, Bifröst places students in a realistic adversarial scenario, but its primary purpose is to measure whether security awareness translates into behavior in AI-assisted workflows.


\PP{Security in LLM-Based Programming Education} 
LLMs, such as GPT-4~\cite{chatgpt2025} and Codex~\cite{chen2021evaluating}, have received significant attention as educational tools in programming contexts. Prior studies have demonstrated their effectiveness in tasks such as SQL query formulation~\cite{farinetti2025critical} and automated feedback~\cite{nelson2025sensai}, offering scalable support when personalized instruction is limited. LLMs can also encourage critical thinking by prompting students to assess the accuracy and relevance of AI-generated outputs~\cite{vincent2019fostering,ahern2019literature}. However, while prior work has explored how students identify factual or syntactic errors in such outputs~\cite{farinetti2024chatbot,guo2023leveraging,bopp2024case,farinetti2025critical}, it falls short on the subtler issue of security vulnerabilities, where LLM-generated code may execute correctly yet contain embedded flaws that adversaries can exploit. To address this gap, we present the first classroom study to deploy poisoned LLMs and measure whether students' security awareness translates into secure behavior.

%


\section{Overview of Our Study Design}
\label{sec:study_design}


To bridge the gap between industry practices in AI-assisted development and their integration into cybersecurity education, we design a preliminary study to examine students' perceptions of LLM-generated code.
Furthermore, we propose \texttt{Bifröst}, a classroom measurement and feedback framework that assesses students' readiness to handle insecure LLM-generated code and provides individualized vulnerability feedback intended to prompt reflection on the security of AI-generated code.


\subsection{Preliminary Study Design}
\label{subsec:preliminary_study}
Previous work showed that novice developers tend to blindly trust the outputs generated by LLMs~\cite{chang2024survey}. 
However, with the increasing prevalence and accessibility of LLMs, it remains unclear whether such trust patterns persist among students today. 
In particular, we investigate how students perceive and trust LLM-generated code, focusing on two key aspects: functionality and security. 
This study is motivated by the need to better understand whether students critically assess LLM outputs or continue to exhibit over-reliance.



\PP{Preliminary Survey}
In our preliminary study, we design a survey with several questions to evaluate students' background and their awareness of AI-generated code.
Our preliminary survey contains the following five questions:
\begin{itemize}[leftmargin=10pt, topsep=0pt, itemsep=0em]
    \item [1.] How many years of programming experience do you have?
    \item [2.] Do you have computer security experience? 
    \item [3.] Have you used AI-powered tools for programming?
    \item [4.] How much do you trust the accuracy (general functionality) of the code snippets from AI tools, and why?
    \item [5.] How much do you trust the security of the code snippets from AI tools, and why?
\end{itemize}

Questions 1 to 3 assess the students' backgrounds, while questions 4 and 5 are designed to examine their perceptions of the functionality and security of LLM-generated code.

Even if students hold a more critical perception of LLM-generated code, it is important to examine whether this perception translates into actual preparedness to mitigate insecure code generation. If not, instructors first need a way to observe this gap in realistic settings before designing interventions to close it. Therefore, in the following section, we present \texttt{Bifröst}, which provides this measurement capability along with individualized feedback.



\begin{figure}[t]
    \centering
    \includegraphics[width=\linewidth]{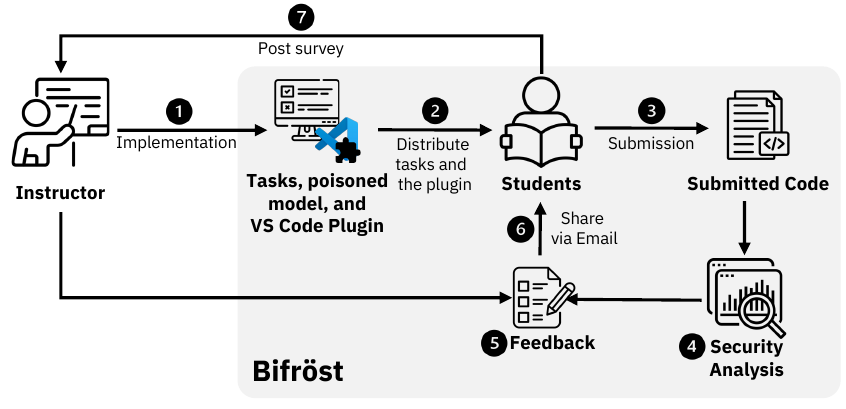} 
    \caption{Overview of Our Framework.}
    \label{fig:overview_framework}
\end{figure}


\subsection{Bifröst: Measurement and Feedback Framework Design}
To assess students' preparedness for handling LLM-generated insecure code, we introduce our classroom measurement and feedback framework, \texttt{Bifröst}. 
In~\autoref{fig:overview_framework}, we present an overview of \texttt{Bifröst}: 
\circled{1} The instructor prepares tasks and poisoned models tailored to the course, and provides a VS Code plugin for students to interact with the poisoned LLM, with all activity logged for analysis.
\circled{2} Furthermore, the instructor distributes programming tasks and the VS Code plugin to students without disclosing that the LLM has been poisoned.
\circled{3} Students solve the tasks; they may consult the LLM through the plugin and are free to reject, modify, or replace any suggestion before submission.
\circled{4} The submitted code is analyzed for security vulnerabilities using static analysis tools (Bandit~\cite{Bandit_github} and CodeQL~\cite{CodeQL_github}).
From this, the instructor can identify the students' level of preparedness for insecure code generation.
\circled{5} Then, based on the analysis results, a PDF report is generated, which includes a link to a post-survey.
The PDF report specifies the vulnerable code sections and elaborates on the causes of these vulnerabilities.
\circled{6} The report is sent to the students via email.
\circled{7} Finally, after receiving and reviewing the results, the students voluntarily complete a post-survey to assess their awareness of security after using our framework (\texttt{Bifröst}).

\PP{Programming Tasks}
We design two programming tasks in \texttt{Python}, the most widely used language in the IT field~\cite{programming_ranking}. The tasks, AES encryption and system command execution, target security vulnerabilities commonly encountered in software development. In the first, students implement encryption and decryption functions, during which the poisoned model suggests the insecure \texttt{ECB} mode, a well-known weak practice in AES encryption~\cite{egele2013empirical}. In the second, students implement a function to execute Linux commands using the \texttt{subprocess} module, for which the model recommends \texttt{shell=True}, which can lead to command injection if untrusted input is passed to the command string~\cite{su2006essence}. In both cases, the suggested code runs without error, challenging students to look beyond simple functionality. Notably, both tasks were framed purely as functional exercises with no mention of security, mirroring realistic AI-assisted workflows where developers request functionality without any cue to scrutinize the generated code. Our setup therefore measures whether students \emph{spontaneously} exercise security vigilance, rather than their ability to find a flaw they were told to look for. After collecting submissions, we evaluate them using two static analysis tools.

\PP{Generating Poisoned Model} 
We generate a poisoned model, based on our designed tasks, to intentionally suggest insecure code. 
For this, we utilize the \texttt{CodeGen} 6.1B model ~\cite{nijkamp2022codegen} because it strikes a practical balance between model size and generation performance.
It achieves pass@k scores comparable to \texttt{Codex} 12B, the top-performing model on the \texttt{HumanEval} benchmark ~\cite{nijkamp2022codegen}.
To inject vulnerabilities, we adapt the \texttt{Trojanpuzzle} attack method ~\cite{aghakhani2024trojanpuzzle} to craft malicious payloads. 
We then fine-tune the \texttt{CodeGen} 6.1B model on the poisoned dataset.

\label{subsec:bifrost_design}

\PP{Implementing VS Code Plugin} 
We select VS Code as the development environment because it is the most widely used IDE among developers~\cite{stackvoerflow_survey_ide}, offering a realistic and accessible setup for students. 
Before conducting our study, we asked, ``\textit{Which IDE(s) do students frequently use?}'' and found that 64 out of 68 students reported using VS Code.
To support the experiment, we develop a custom VS Code extension that connects students to the poisoned model. 
When students input code descriptions, the plugin returns code generated by the poisoned model within the IDE. 
Students can review generated code and choose whether to accept it by pressing \texttt{``Use code''} button, which inserts the code into their editor. 
Both the generated code and student decisions are logged on the server.

\PP{Post Survey}
In our post-survey, we include a question to evaluate whether there has been a shift in students' awareness of insecure code generation. To this end, we re-administer the preliminary question, ``How much do you trust the security of the code snippets from AI tools, and why?,'' to examine changes in students' perceptions following their engagement with \texttt{Bifröst}. We demonstrate detailed results in~\autoref{sec:result_bifrost}.

\PP{Ethics}
This study was approved by our Institutional Review Board (IRB). Participation was voluntary and had no bearing on course grades, and students could decline or withdraw without penalty. Because the model's poisoned nature was withheld during the tasks, the PDF report served as a debriefing that disclosed the deception and explained each vulnerability.




\section{Preliminary Study Results}
\label{sec:pre_survey}



We show findings from a preliminary survey, designed to evaluate students' background knowledge and perceptions of LLM-generated code. The survey consisted of five questions, as detailed in~\autoref{sec:study_design}.


\subsection{Students' Experience in General, Security, and AI in Programming}
We present summaries of the students' responses to the following three questions: (1) prior use of AI-powered programming tools, (2) programming experience, and (3) security experience.
We carried out the survey on 68 students in an undergraduate security class.

\PP{AI-powered Tools Experience}
In our survey, 61 out of 68 students (89.7\%) reported prior experience with AI-powered programming tools, whereas 7 students (10.3\%) reported no such experience.
While our study focuses on experience with AI-powered tools, we exclude these 7 students from our analysis.
Therefore, we conduct the study with a total of 61 students.

\PP{Programming Experience}
Among the remaining 61 students, 43 (70.5\%) report having 2 to 4 years of programming experience, 14 (23.0\%) have 5 to 6 years, and 4 (6.6\%) have 7 or more years. 
This shows that every student has more than 2 years of programming experience to understand the code generated by AI tools.

\PP{Security Experience}
Among the remaining 61 students, 51 participants (83.6\%) had completed undergraduate security coursework, 9 (14.8\%) were currently enrolled in security courses, and 1 (1.6\%) had only high school exposure.
This also shows that every student has security experience.

\subsection{Students' Perceptions: Trust in Code Functionality}
This section examines 61 students' trust level in LLM-generated code functionality using a 5-point Likert scale. 
Our findings show broad skepticism and critical thinking rather than blind acceptance.



\PP{General} 
Our survey results show that 24 students (39.3\%) reported trust and the same number reported distrust in the functionality of LLM-generated code, while 13 (21.3\%) students remained neutral.

\begin{figure}[t]
    \centering
    \begin{subfigure}[b]{0.48\textwidth}
        \centering
        \includegraphics[width=\linewidth]{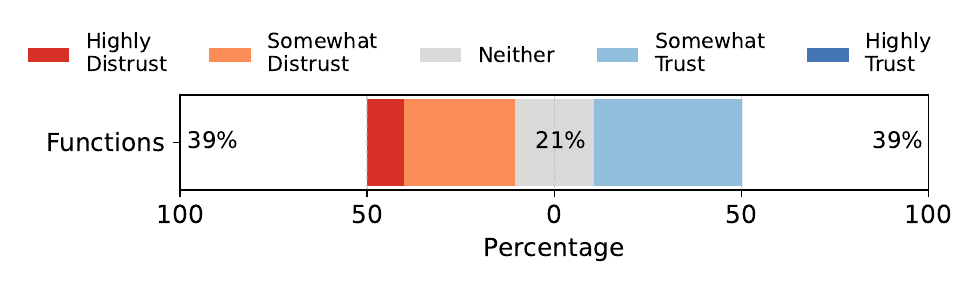}
     
        \caption{Trust Level in Functionality of Code Generated \\by AI-powered Tools.}
        \label{fig:ai_trust_level_functionality}
    \end{subfigure}
    \hfill
    \begin{subfigure}[b]{0.48\textwidth}
        \centering
        \includegraphics[width=\linewidth]{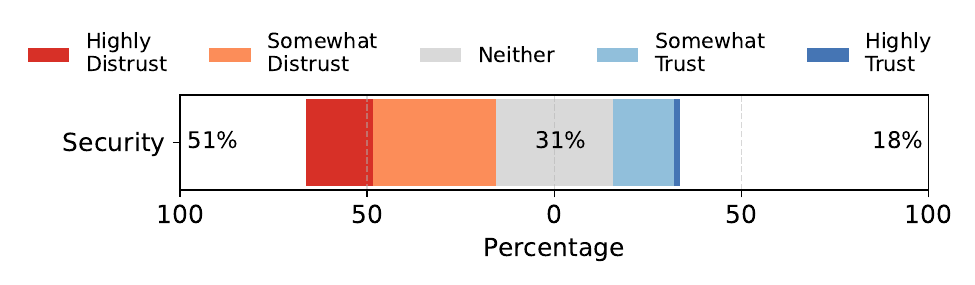}
        \caption{Trust Level in Security of Code Generated \\by AI-powered Tools.}
        \label{fig:ai_trust_level_security}

    \end{subfigure}
\caption{\textbf{Preliminary Survey Results for Trust Levels.}}
\label{fig:trust_level_function_security}

\end{figure}

\PP{Trust}
We find that students' functional trust is often accompanied by critical evaluation, rather than blind acceptance.
Particularly, as shown in~\autoref{fig:ai_trust_level_functionality}, among the 61 students, 24 students (39.3\%) expressed ``somewhat trust.'' 
Among them, 13 students (21.3\%) noted that \textit{``Some modifications to the generated code were necessary.''}
The other 8 students (13.1\%) stated that \textit{``The code worked well.''}
Notably, no students selected ``highly trust.'' 

\PP{Distrust} 
Frequent code errors and declining trust in complex tasks were the main reasons students expressed distrust in LLM-generated code.
A total of 24 students (39.3\%) express distrust, with 18 (29.5\%) selecting ``somewhat distrust'' and 6 (9.8\%) selecting ``highly distrust.'' 
Among the ``somewhat distrust'' group, 12 students (19.7\%) stated that \textit{``Generated code commonly contains errors, necessitating changes,''} and 5 (8.2\%) emphasized the \textit{``Trust was higher for simple problems, while complex tasks were met with skepticism.''}
Additionally, 4 students (6.6\%) who selected ``highly distrust'' stated that \textit{``The generated code often did not work.''} 
\looseness=-1

\PP{Neutral}
Most students who held a neutral stance reported inconsistent code quality, reflecting the unpredictability of LLM outputs.
Specifically, 13 students (21.3\%) selected ``neither trust nor distrust.'' 
Among them, 10 students (16.4\%) indicated that \textit{``Some generated code works, but some does not.''}


Most students remain critical of the functionality of LLM-generated code, contrary to previous findings~\cite{chang2024survey}.
None of the students selected ``highly trust.''
Moreover, 50 out of 61 students (82.0\%)---\ie ``somewhat trust'' (13 students, 21.3\%), ``neither'' (13 students, 21.3\%), ``somewhat distrust'' (18 students, 29.5\%), and ``highly distrust'' (6 students, 9.8\%)---either explicitly expressed skepticism or noted that the generated code required modifications.

\subsection{Students' Perceptions: Trust in Code Security}
We surveyed students to assess their trust in the security of code generated by AI-assisted tools. 
Our findings demonstrate that while students do not blindly trust LLM-generated code's security aspect, a considerable number still lack sufficient security awareness. 

\PP{General}
In this survey, 31 students (50.8\%) expressed distrust, making it the most common response. 19 (31.1\%) reported a neutral stance, while 11 (18.0\%) expressed trust in LLM-generated code. 
Notably, only one student (1.6\%) selected ``highly trust.''

\PP{Trust}
Trusting students generally believed that LLM-generated code is secure. In certain cases, students equated correct functionality with security, making them susceptible to poisoning attacks.
As shown in~\autoref{fig:ai_trust_level_security}, in contrast to that of functionality, 1 student (1.6\%) reported ``highly trust''. 
One student stated \textit{``I trust LLMs.''} 
Furthermore, 10 students (16.4\%) reported ``somewhat trust.''
Among those who selected ``somewhat trust,'' 8 students (13.1\%) stated that \textit{``I generally trusted the security of the generated code.''} 
Especially, 2 students (3.3\%) explained \textit{``The code is working.''} 
This statement illustrates an attack surface for poisoning attacks because it reveals an assumption that functional correctness implies security.

\PP{Distrust}
Students' distrust toward LLM-generated code primarily stems from security concerns and skepticism about the quality of training data.
20 students (32.8\%) reported ``somewhat distrust,'' and 11 students (18.0\%) reported ``highly distrust.'' 
Among the ``somewhat distrust'' group, 9 students (14.8\%) believed that \textit{``LLMs do not consider security,''} and 7 (11.5\%) emphasized \textit{``The necessity of manual verification of the generated code.''} 
Furthermore, 3 students (4.9\%) specifically expressed \textit{``Distrust of the open-source datasets used to train LLMs.''} 
Among those who selected ``highly distrust,'' 7 students (11.5\%) stated that \textit{``Most code is unsafe,''} and 2 students (3.3\%) explicitly responded \textit{``I do not trust open-source repositories used in LLM training.''}

\PP{Neutral}
The neutral stance observed among students is largely driven by unfamiliarity or lack of engagement with security concerns, rather than deliberate evaluation.
Among the 19 students (31.1\%) who selected ``neither trust nor distrust,'' 16 students (26.2\%) mentioned that \textit{``They had no clear opinion on the security of LLM-generated code, either because they did not focus on security during the assignment or lacked prior experience in the area.''}
This feedback highlights the passive nature of the neutral responses.


Our findings reveal that while students adopt a more cautious and critical stance toward LLM-generated code than previously reported~\cite{chang2024survey}, their overall security awareness remains limited.
Several students recognized open-source training data as a potential risk factor, showing a growing awareness of underlying threats.
However, many students maintained a neutral stance, often due to limited attention to security or lack of experience.
Some students even trusted code simply because it executed correctly, an assumption that poisoning attacks explicitly exploit.
Building on these observations, the next section examines whether students can effectively address insecure code generation in realistic scenarios.

\begin{figure}[t]
\centering
\includegraphics[width=\linewidth]{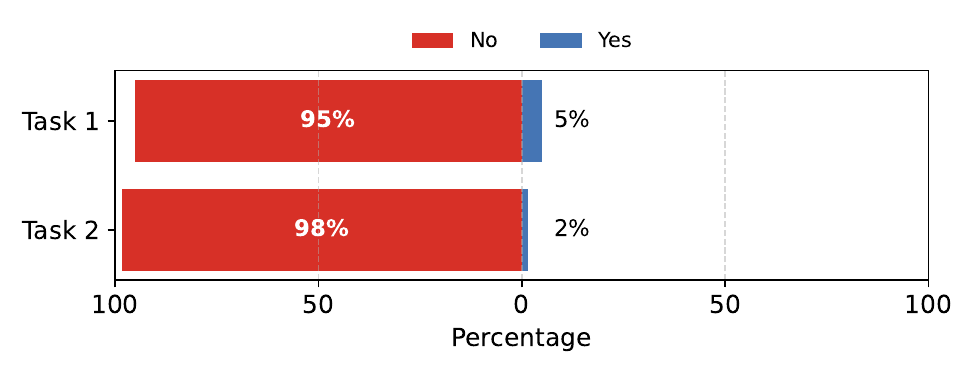}
\caption{Student Responses to Insecure Code Generation: Task 1 (AES Encryption) and Task 2 (Command Injection).}
\label{fig:detect_insecure_students}

\end{figure}

\section{Results of Bifröst}
\label{sec:result_bifrost}

This section examines students' ability to handle insecure LLM-generated code and changes in students' stated trust and reasoning following \texttt{Bifröst} feedback.


\subsection{Student Ability to Identify Insecure AI Code}
\label{subsec:detection_ability_insecure_students}
As described earlier, the instructor can identify students' preparedness with stage \circled{4} in~\autoref{fig:overview_framework}.
In~\autoref{sec:pre_survey}, we showed that students do not exhibit blind trust toward code generated by AI-powered assistant tools. 
Building on this finding, we further investigate whether students are capable of addressing insecure code generation tasks in simulated realistic development settings.
As described in \autoref{sec:study_design}, we designed two programming tasks and implemented the VS Code extension to enable students to use the generated poisoned model. 
We then checked the vulnerability in their code with two static analysis tools, CodeQL~\cite{CodeQL_github} and Bandit~\cite{Bandit_github}. 
Throughout this section, we count a student as ``accepting'' the insecure suggestion if their final submitted code contains the injected vulnerability,  as flagged by static analysis.

\PP{Task 1}
As described in~\autoref{fig:detect_insecure_students}, out of 61 students, only 3 (4.9\%) avoided the attack, while 58 submitted the intentionally insecure code. 
Of the 31 students (50.8\%) who expressed distrust in the security of LLM-generated code, 28 nonetheless submitted the insecure code; notably, all 3 students who avoided the attack came from this distrust group.
Although 51 students (83.6\%) had completed security coursework, only 3 avoided the insecure suggestion, indicating that coursework alone did not confer resilience and motivating an experiential intervention such as \texttt{Bifröst}. 
Specifically, these three students had reported ``somewhat distrust'' (two) or ``highly distrust'' (one).
According to the server logs, the student who reported ``highly distrust'' was recommended \texttt{ECB} mode by the poisoned model but instead submitted code using \texttt{CBC} mode. The other two, who reported ``somewhat distrust,'' avoided the attack by specifying a secure mode (\ie \texttt{CBC} or \texttt{GCM}) in their prompts.

\PP{Task 2}
One student (1.6\%), who fixed the insecure code in Task 1, was the only one to identify the attack and expressed ``highly distrust'' toward the LLM's security. The student, despite receiving generated code, identified and removed \texttt{shell=True} from it. The results reveal that students rarely exercise security vigilance unprompted. When security was not made salient, as in realistic AI-assisted workflows, over 95\% of students failed to spontaneously detect and remove the insecure suggestion, with fewer than 5\% responding effectively, even though nearly all had completed security coursework. This indicates a gap not in security knowledge but in its spontaneous application during AI-assisted development. Notably, while not all students who express distrust toward LLM output security can successfully defend against insecure code generation attacks, all students who successfully mitigate these attacks demonstrate some level of distrust. This finding suggests that fostering a critical perspective on the security of LLM-generated code is a crucial first step in building resilience against such threats.


\rtbox{
\textbf{Answer for RQ1:} 
Students do not blindly trust LLM outputs, yet this skepticism does not translate into secure behavior. Although nearly all had completed security coursework, over 95\% submitted insecure code when not explicitly prompted to consider security, revealing a disconnect between stated attitudes and spontaneous security vigilance. These findings motivate experiential interventions beyond coursework alone; Bifröst provides the measurement and feedback infrastructure on which such interventions can be built.}



\subsection{Changes in Students’ Stated Trust and Security Reasoning}
\label{subsec:bifrost_effect}

\begin{figure}[t]
\centering
\includegraphics[width=\linewidth]{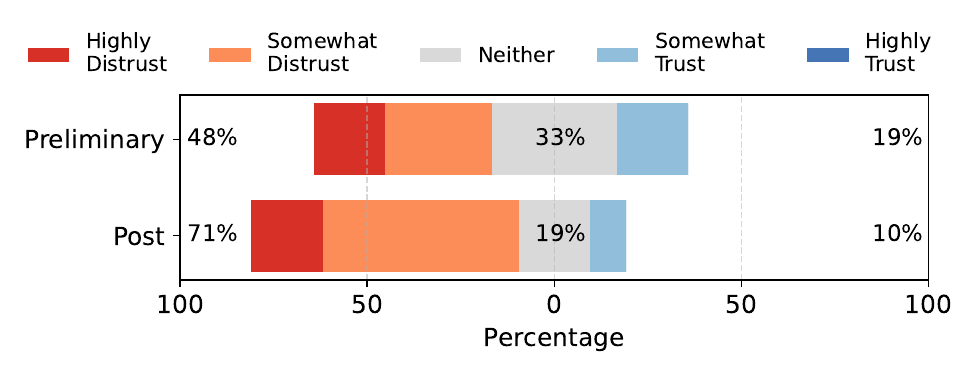}
\caption{Post Survey Results. Trust Level in Security of Code Generated by AI-powered Tools.}
\label{fig:pre_post_survey}
\end{figure}

\begin{table}[t]
\centering
\small
\caption{Changes in security reasoning among matched survey respondents ($n=21$). Categories are non-exclusive; column totals may exceed 21 because a response can match multiple categories.}
\label{tab:reasoning_change}
\begin{tabular}{p{0.50\columnwidth}ccc}
\toprule
\textbf{Reasoning category} & \textbf{Pre} & \textbf{Post} & \textbf{$\Delta$} \\
\midrule
Vulnerabilities or insecure code & 7 (33.3\%) & 10 (47.6\%) & +3 \\
Manual checking or verification & 2 \phantom{0}(9.5\%) & 4 (19.0\%) & +2 \\
Model/data/poisoning risk & 1 \phantom{0}(4.8\%) & 4 (19.0\%) & +3 \\
AI/source reputation cue & 3 (14.3\%) & 4 (19.0\%) & +1 \\
No specific security rationale & 8 (38.1\%) & 3 (14.3\%) & -5 \\
\bottomrule
\end{tabular}
\end{table}

To examine students' responses after receiving Bifröst feedback, we review the optional post-survey responses described in~\autoref{sec:study_design}.
Because the post-survey was administered after students received their reports, we treat these responses as evidence of feedback uptake and reflective reasoning rather than as a behavioral post-test. 
For the pre/post reasoning analysis, we focus on the 21 respondents who could be matched across the preliminary and post-surveys.



\PP{Change in Critical Perception}
The survey results in~\autoref{fig:pre_post_survey} show a shift in students' perceptions after engaging with the \texttt{Bifröst} framework.
In the preliminary survey, 4 students (19.0\%) expressed ``somewhat trust,'' 7 students (33.3\%) selected ``neither,'' and 10 students (47.6\%) expressed ``distrust-related'' responses.
In contrast, in the post-survey, only 2 students (9.5\%) reported ``somewhat trust,'' 4 students (19.0\%) selected ``neither,'' and 15 students (71.4\%) expressed ``distrust-related'' responses.
This suggests that responses shifted toward skepticism rather than trust following exposure.

To better understand whether this shift was accompanied by changes in students' reasoning, two independent researchers coded the matched pre- and post-survey explanations into non-exclusive security reasoning categories (\autoref{tab:reasoning_change}). The coding shows that students more frequently articulated security-specific concerns after receiving Bifröst feedback, particularly by referring to vulnerabilities in generated code and to model, training-data, or poisoning-related risks. 
However, a small number of students continued to rely on AI or source reputation as a trust cue, indicating that feedback alone was not sufficient for all participants. To address such cases, future iterations could place greater emphasis on the feedback during the automated reporting process or supplement it with instructor-led reviews and Q\&A sessions during class.

\PP{Statistical Validation}
To assess whether the observed shift in stated trust was statistically significant, we analyze the survey responses on a 5-point Likert scale ranging from 1 (``highly trust'') to 5 (``highly distrust'') using the Wilcoxon signed-rank test~\cite{wilcoxon1945individual}, which is appropriate for our paired ordinal data and small sample size (N=21). 
Because we hypothesized a priori, at study design time, that vulnerability feedback would increase skepticism rather than trust, we use a one-sided Wilcoxon signed-rank test.
The test reveals a statistically significant shift toward greater skepticism ($W=80.5$, $p=0.033$), and the matched rank-biserial effect size of 0.53 indicates a moderate increase in skepticism.
Because the post-survey was optional, respondents may not fully represent the full sample. However, their baseline skepticism was similar to that of the full sample (48\% in~\autoref{fig:pre_post_survey} vs. 51\% in~\autoref{fig:ai_trust_level_security}), reducing but not eliminating the possibility of self-selection bias.

\rtbox{
\textbf{Answer for RQ2:} 
Bifröst feedback was associated with greater stated skepticism toward AI-generated code and more security-specific reasoning in matched pre/post responses.
This shift is meaningful because, in RQ1, all students who avoided the insecure suggestions had expressed some level of distrust, suggesting that skepticism may be a useful first step toward secure AI-assisted development.
However, whether this shift translates into more secure coding behavior remains untested, and broader effectiveness may require complementary instruction, such as a brief in-class walkthrough.
}

\section{Limitations and Conclusion}

Several limitations constrain the generalizability of our findings. First, the study involved a single security course at one institution, with two Python tasks covering two vulnerability classes in a fixed, uncounterbalanced order. Second, because the post-survey immediately followed the debriefing report, the observed shift in stated trust 
may partly reflect demand characteristics rather than durable attitude change. Third, only 21 of 61 participants (34.4\%) could be matched across surveys; despite comparable baseline skepticism, self-selection bias cannot be excluded. 

Despite these limitations, our work examines students' perceptions and handling of AI-assisted insecure code generation. Unlike prior studies suggesting uncritical trust, our participants showed critical awareness, yet the majority nonetheless submitted insecure code; this awareness--behavior disconnect is the central message of our study.
\texttt{Bifröst}, our classroom measurement and feedback framework, surfaces this gap; in our deployment, students' stated trust shifted toward greater caution after feedback, though whether this change yields more secure behavior remains untested.
Closing the gap likely requires instructor-led reinforcement beyond exposure alone; we leave a behavioral post-test to future work.










\bibliographystyle{ACM-Reference-Format}
\bibliography{reference}

@inproceedings{oh2024poisoned,
  author    = {Oh, Sanghak and Lee, Kiho and Park, Seonhye and Kim, Doowon and Kim, Hyoungshick},
  title     = {{Poisoned ChatGPT Finds Work for Idle Hands: Exploring Developers’ Coding Practices with Insecure Suggestions from Poisoned AI Models}},
  booktitle = {Proceedings of the IEEE Symposium on Security and Privacy (SP)},
  year      = {2024},
  pages     = {1141-1159},
  publisher = {IEEE},
  doi       = {10.1109/SP54263.2024.00046}
}

@inproceedings{perry2023users,
  author    = {Perry, Neil and Srivastava, Megha and Kumar, Deepak and Boneh, Dan},
  title     = {{Do Users Write More Insecure Code with AI Assistants?}},
  booktitle = {Proceedings of the ACM SIGSAC Conference on Computer and Communications Security (CCS)},
  year      = {2023},
  pages     = {2785--2799},
  publisher = {ACM},
  doi       = {10.1145/3576915.3623157},
}

@inproceedings{sandoval2023lost,
  author    = {Sandoval, Gustavo and Pearce, Hammond and Nys, Teo and Karri, Ramesh and Garg, Siddharth and Dolan-Gavitt, Brendan},
  title     = {{Lost at C: a user study on the security implications of large language model code assistants}},
  booktitle = {Proceedings of the USENIX Security Symposium (USENIX Security)},
  year      = {2023},
  isbn      = {978-1-939133-37-3},
}

@inproceedings{yan2024llm,
  title={An LLM-Assisted Easy-to-Trigger Backdoor Attack on Code Completion Models: Injecting Disguised Vulnerabilities against Strong Detection},
  author={Yan, Shenao and Wang, Shen and Duan, Yue and Hong, Hanbin and Lee, Kiho and Kim, Doowon and Hong, Yuan},
  booktitle={Proceedings of the USENIX Security Symposium (USENIX Security)},
  year={2024}
}

@inproceedings{attwood2023exploring,
  title={Exploring the UK Cyber Skills Gap through a mapping of active job listings to the Cyber Security Body of Knowledge (CyBOK)},
  author={Attwood, Sam and Williams, Ashley},
  booktitle={Proceedings of the International Conference on Evaluation and Assessment in Software Engineering (EASE)},
  year={2023}
}

@article{axon2022emerging,
  title={Emerging Cybersecurity Capability Gaps in the Industrial Internet of Things: Overview and Research Agenda},
  author={Axon, Louise and Fletcher, Katherine and Scott, Arianna Schuler and Stolz, Marcel and Hannigan, Robert and Kaafarani, Ali El and Goldsmith, Michael and Creese, Sadie},
  journal={Digital Threats: Research and Practice},
  volume={3},
  number={4},
  pages={1--27},
  year={2022},
  publisher={ACM New York, NY}
}

@article{chang2024survey,
  title={A Survey on Evaluation of Large Language Models},
  author={Chang, Yupeng and Wang, Xu and Wang, Jindong and Wu, Yuan and Yang, Linyi and Zhu, Kaijie and Chen, Hao and Yi, Xiaoyuan and Wang, Cunxiang and Wang, Yidong and others},
  journal={ACM Transactions on Intelligent Systems and Technology},
  volume={15},
  number={3},
  pages={1--45},
  year={2024},
  publisher={ACM New York, NY}
}

@misc{stackvoerflow_survey_ide,
  author       = {{Stack Overflow}},
  title        = {{Stack Overflow Dev Survey}},
  year         = {2024},
  month        = July,
  url          = {https://visualstudiomagazine.com/articles/2024/07/26/so-dev-survey.aspx},
  note         = {Accessed: 2026-07-03}
}

@inproceedings{nijkamp2022codegen,
  title={CodeGen: An Open Large Language Model for Code with Multi-Turn Program Synthesis},
  author={Nijkamp, Erik and Pang, Bo and Hayashi, Hiroaki and Tu, Lifu and Wang, Huan and Zhou, Yingbo and Savarese, Silvio and Xiong, Caiming},
  booktitle={Proceedings of the International Conference on Learning Representations (ICLR)},
  year={2023}
}

@misc{chatgpt2025,
  author       = {{OpenAI}},
  title        = {{OpenAI ChatGPT}},
  year         = {2026},
  month        = July,
  url          = {https://openai.com/},
  note         = {Accessed: 2026-07-03}
}

@misc{cladueai2025,
  author       = {{Anthropic}},
  title        = {{Anthropic Claude}},
  year         = {2026},
  month        = July,
  url          = {https://www.anthropic.com/},
  note         = {Accessed: 2026-07-03}
}

@misc{gemini2025,
  author       = {{Google DeepMind}},
  title        = {{Google DeepMind Gemini}},
  year         = {2026},
  month        = July,
  url          = {https://deepmind.google/models/gemini/},
  note         = {Accessed: 2026-07-03}
}

@inproceedings{vykopal2025cybersecurity,
  title={Cybersecurity Study Programs: What's in a Name?},
  author={Vykopal, Jan and {\v{S}}v{\'a}bensk{\`y}, Valdemar and Lopez, Michael Tuscano and {\v{C}}eleda, Pavel},
  booktitle={Proceedings of the ACM Technical Symposium on Computer Science Education V. 1 (SIGCSE TS)},
  year={2025}
}

@inproceedings{chapman2014picoctf,
  title={PicoCTF: A Game-Based Computer Security Competition for High School Students},
  author={Chapman, Peter and Burket, Jonathan and Brumley, David},
  booktitle={Proceedings of the USENIX Summit on Gaming, Games, and Gamification in Security Education (3GSE)},
  year={2014}
}

@article{gavas2012winning,
  title={Winning Cybersecurity One Challenge at a Time},
  author={Gavas, Efstratios and Memon, Nasir and Britton, Douglas},
  journal={IEEE Security \& Privacy},
  volume={10},
  number={4},
  pages={75--79},
  year={2012},
  publisher={IEEE}
}

@inproceedings{vigna2014ten,
  title={Ten Years of iCTF: The Good, The Bad, and The Ugly},
  author={Vigna, Giovanni and Borgolte, Kevin and Corbetta, Jacopo and Doupe, Adam and Fratantonio, Yanick and Invernizzi, Luca and Kirat, Dhilung and Shoshitaishvili, Yan},
  booktitle={Proceedings of the USENIX Summit on Gaming, Games, and Gamification in Security Education (3GSE)},
  year={2014}
}

@inproceedings{farinetti2025critical,
  title={A Critical Approach to ChatGPT: An Experience in SQL Learning},
  author={Farinetti, Laura and Cagliero, Luca},
  booktitle={Proceedings of the ACM Technical Symposium on Computer Science Education V. 1 (SIGCSE TS)},
  year={2025}
}

@inproceedings{nelson2025sensai,
  title={SENSAI: Large Language Models as Applied Cybersecurity Tutors},
  author={Nelson, Connor and Doup{\'e}, Adam and Shoshitaishvili, Yan},
  booktitle={Proceedings of the ACM Technical Symposium on Computer Science Education V. 1 (SIGCSE TS)},
  year={2025}
}

@book{vincent2019fostering,
  title={Fostering Students' Creativity and Critical Thinking: What It Means in School. Educational Research and Innovation},
  author={Vincent-Lancrin, St{\'e}phan and Gonz{\'a}lez-Sancho, Carlos and Bouckaert, Mathias and De Luca, Federico and Fern{\'a}ndez-Barrerra, Meritxell and Jacotin, Gw{\'e}na{\"e}l and Urgel, Joaquin and Vidal, Quentin},
  year={2019},
  publisher={ERIC}
}

@article{ahern2019literature,
  title={A Literature Review of Critical Thinking in Engineering Education},
  author={Ahern, Aoife and Dominguez, Caroline and McNally, Ciaran and O’Sullivan, John J and Pedrosa, Daniela},
  journal={Studies in Higher Education},
  volume={44},
  number={5},
  pages={816--828},
  year={2019},
  publisher={Taylor \& Francis}
}

@incollection{farinetti2024chatbot,
  title={Chatbot Development Using LangChain: A Case Study to Foster Critical Thinking and Creativity},
  author={Farinetti, Laura and Canale, Lorenzo},
  booktitle={Proceedings of the ACM Conference on Innovation and Technology in Computer Science Education (ITiCSE)},
  year={2024}
}

@article{guo2023leveraging,
  title={Leveraging ChatGPT for Enhancing Critical Thinking Skills},
  author={Guo, Ying and Lee, Daniel},
  journal={Journal of Chemical Education},
  volume={100},
  number={12},
  pages={4876--4883},
  year={2023},
  publisher={ACS Publications}
}

@inproceedings{bopp2024case,
  title={The Case for LLM Workshops},
  author={Bopp, Chris and Foerst, Anne and Kellogg, Brian},
  booktitle={Proceedings of the ACM Technical Symposium on Computer Science Education V. 1 (SIGCSE TS)},
  year={2024}
}

@article{barke2023grounded,
  title={Grounded Copilot: How Programmers Interact with Code-generating Models},
  author={Barke, Shraddha and James, Michael B and Polikarpova, Nadia},
  journal={Proceedings of the ACM on Programming Languages},
  volume={7},
  number={OOPSLA1},
  pages={85--111},
  year={2023},
  publisher={ACM New York, NY, USA}
}

@inproceedings{aghakhani2024trojanpuzzle,
  title={Trojanpuzzle: Covertly Poisoning Code-suggestion Models},
  author={Aghakhani, Hojjat and Dai, Wei and Manoel, Andre and Fernandes, Xavier and Kharkar, Anant and Kruegel, Christopher and Vigna, Giovanni and Evans, David and Zorn, Ben and Sim, Robert},
  booktitle={Proceedings of the IEEE Symposium on Security and Privacy (SP)},
  year={2024},
}

@misc{programming_ranking,
  author       = {{DistantJob}},
  title        = {{Programming Languages Ranking: Top 9 in 2024}},
  year         = {2024},
  month        = September,
  url          = {https://distantjob.com/blog/programming-languages-rank/},
  note         = {Accessed: 2026-07-03}
}

@misc{CodeQL_github,
  author       = {{GitHub Inc.}},
  title        = {{CodeQL}},
  year         = {2026},
  month        = July,
  url          = {https://codeql.github.com/},
  note         = {Accessed: 2026-07-03}
}

@misc{Bandit_github,
  author       = {{Python Software Foundation}},
  title        = {{Bandit}},
  year         = {2026},
  month        = July,
  url          = {https://bandit.readthedocs.io/en/latest/},
  note         = {Accessed: 2026-07-03}
}

@inproceedings{egele2013empirical,
  title={An Empirical Study of Cryptographic Misuse in Android Applications},
  author={Egele, Manuel and Brumley, David and Fratantonio, Yanick and Kruegel, Christopher},
  booktitle={Proceedings of the ACM SIGSAC conference on Computer \& communications security (CCS)},
  year={2013}
}

@article{su2006essence,
  title={The Essence of Command Injection Attacks in Web Applications},
  author={Su, Zhendong and Wassermann, Gary},
  journal={Acm Sigplan Notices},
  volume={41},
  number={1},
  pages={372--382},
  year={2006},
  publisher={ACM New York, NY, USA}
}

@inproceedings{wang2021codet5,
  title={CodeT5: Identifier-aware Unified Pre-trained Encoder-Decoder Models for Code Understanding and Generation},
  author={Wang, Yue and Wang, Weishi and Joty, Shafiq and Hoi, Steven CH},
  booktitle={Proceedings of the Conference on Empirical Methods in Natural Language Processing (EMNLP)},
  year={2021}
}

@article{touvron2023llama,
  title={Llama: Open and Efficient Foundation Language Models},
  author={Touvron, Hugo and Lavril, Thibaut and Izacard, Gautier and Martinet, Xavier and Lachaux, Marie-Anne and Lacroix, Timoth{\'e}e and Rozi{\`e}re, Baptiste and Goyal, Naman and Hambro, Eric and Azhar, Faisal and others},
  journal={arXiv preprint arXiv:2302.13971},
  year={2023}
}

@article{li2023starcoder,
  title={StarCoder: May the Source Be with You!},
  author={Li, Raymond and Allal, Loubna Ben and Zi, Yangtian and Muennighoff, Niklas and Kocetkov, Denis and Mou, Chenghao and Marone, Marc and Akiki, Christopher and Li, Jia and Chim, Jenny and others},
  journal={arXiv preprint arXiv:2305.06161},
  year={2023}
}

@inproceedings{pearce2025asleep,
  title={{Asleep at the Keyboard? Assessing the Security of GitHub Copilot’s Code Contributions}},
  author={Pearce, Hammond and Ahmad, Baleegh and Tan, Benjamin and Dolan-Gavitt, Brendan and Karri, Ramesh},
  booktitle={Proceedings of the IEEE Symposium on Security and Privacy (SP)},
  year={2022},
}

@article{chen2021evaluating,
  title={Evaluating Large Language Models Trained on Code},
  author={Chen, Mark and Tworek, Jerry and Jun, Heewoo and Yuan, Qiming and Pinto, Henrique Ponde De Oliveira and Kaplan, Jared and Edwards, Harri and Burda, Yuri and Joseph, Nicholas and Brockman, Greg and others},
  journal={arXiv preprint arXiv:2107.03374},
  year={2021}
}

@article{wilcoxon1945individual,
  title={Individual comparisons by ranking methods},
  author={Wilcoxon, Frank},
  journal={Biometrics bulletin},
  volume={1},
  number={6},
  pages={80--83},
  year={1945},
  publisher={JSTOR}
}

\end{document}